# THE ROLE OF TRANSVERSE ANISOTROPIC ELASTIC WAVES AND EHRENFEST RELATIONSHIPS IN THE SUPERCONDUCTING STATE OF THE COMPOUND $Sr_2RuO_4$

P. Contreras
Department of Physics, University of the Andes, 5101, Venezuela.

### ABSTRACT

In this short review, some bulk threshold phenomena of the unconventional superconductivity in the strontium ruthenate compound ($Sr_2RuO_4$) are mentioned. Particularly, in $Sr_2RuO_4$, the Ehrenfest relations, the velocity propagation of the elastic field modes, and the ultrasound attenuation find a common point to study elastic anisotropic lattice field effects at the transition temperature ($T_c$) between the normal and the superconducting state. Ehrenfest relations are suitable for establishing a phase diagram of the $T_c$ point according to symmetry consideration of the electrons pairs in unconventional superconductors. Further studies in this direction could turn into a new role for the spin polarization of the transverse phonon field and its interaction with the conduction electrons in anisotropic crystals with several conducting/superconducting bands such as $Sr_2RuO_4$.

**Keywords**: spin value, phonons dispersion law, elastic transversal waves, Ehrenfest relations, ultrasound attenuation, elastic polarization, triplet pairs, tetragonal crystals, threshold phenomena.

## 1. INTRODUCTION

The physical kinetics (PK) or physics of the transport properties is defined as the science that studies the macro and microscopic transport processes by quasiparticles in crystals that are out of thermodynamic equilibrium (Lifshitz and Pitaevskii, 1987). The PK is a difficult subject due to its complicated mathematical formalism through the collision integral or the Green functions technique. However, the experimental study of low-temperature kinetic transport properties and some threshold phenomena such as the ultrasound attenuation and electronic/phonon heat conduction (Pokrovskii, 1961, Kulik, 1962) helps to clarify several physical aspects relevant to unconventional and multiband superconductivity.

For example, the experimental and theoretical study of the directional ultrasound attenuation below the superconducting temperature ($T_c$) helps to predict the symmetry properties of the pair wave function, and the visualization of the nodes in anisotropic superconductors (Pokrovskii and Toponogov, 1961). Additionally, the electronic thermal conductivity accounts for the physics behavior in superconductors at ultra-low temperatures, such as the relationship between nonmagnetic disorder and superconductivity, allowing to predict whether the superconductor crystal is in Bohr, intermediate or unitary scattering limit as $T$ approaches absolute zero. This effect is called the universal behavior of the kinetic coefficients as the temperature approaches zero value (Griffin, 1965; Ambegaokar and Griffin, 1965; Hirschfeld et al., 1988).

Finally, the Ehrenfest thermodynamic relations (ETR), namely those that link the thermal expansion, the specific heat, and the isothermal compressibility (see table 1) are powerful relationships in the study of second-order phase transitions at $T_c$ (Landau and Lifshitz, 1980; Rumer and Ryvkin, 1980).

---

Corresponding author e-mail: pecontreras@ula.ve

ETR occurs for example, in the second-order transition from a normal metallic state to a superconducting or ferromagnetic/antiferromagnetic ordered phase. Ehrenfest relations are suitable for $4^{th}$ rank elastic tensors as was shown in the study of spin density waves in Chromium (Walker, 1980). In addition, ETR allows to sketch a phase diagram based on the symmetry properties of the pair wave function and discern between the dimension of the order parameter in the superconducting state.

Therefore, in the discussion section of this note, we will make a brief review of results obtained thanks to the analysis of the ETR and the ultrasound attenuation in the bulk of the compound $Sr_2RuO_4$ for a particular superconducting phenomenological model (Walker and Contreras, 2002; Contreras *et al.*, 2004; Contreras, 2011; Contreras *et al.*, 2016). Recently, novel experimental and theoretical advances continue to be achieved in the prediction of the broken time-reversal symmetry state of $Sr_2RuO_4$. These studies are useful in order to explain the microscopic mechanism inherent to superconductivity in this compound (Benhabib *et al.*, 2020; Ghosh *et al.*, 2020).

In the additional remarks section, it is proposed to carry out certain theoretical studies related to three threshold phenomena of interest. First, the propagation of the transverse elastic phonons and their spin value and polarization direction in the normal state of $Sr_2RuO_4$, whether the z-projection of the of the transverse elastic phonons spin in a tetragonal crystal is equal to ±1, since in principle for anisotropic crystals, the spin only will be well defined in certain directions of propagation. Second, the role of the elastic polarization modes for the different sheets of the FS, and third, the changes in sound speed on the dispersion phonon curve at $T_c$.

These studies could give unexpected results inherent to the propagation, polarization, and attenuation of transverse elastic waves by conduction electrons, in compounds with an anisotropic lattice, and several sheets on the FS, such as the crystal for strontium ruthenate. Let's discuss the subject.

## 2. DISCUSSION

The strontium ruthenate crystal ($Sr_2RuO_4$) has a body-centered tetragonal lattice ($D_{4h}$) with a layered square structure for the ruthenium atoms, Figure 1. Its normal state, it is described by a Fermi liquid, with three metallic conduction sheets. Therefore, the Fermi surface (FS) is composed of three sheets called the **$\alpha$, $\beta$**, and **$\gamma$** sheets (C. Bergemann, A. P. Mackenzie, S. R. Julian, D. Forsythe & E. Ohmichi, 2003).

In addition, $Sr_2RuO_4$ is an unconventional superconductor with $T_c \approx 1.5$ K. $T_c$ that strongly depends on the non-magnetic disorder in the crystal sample. $Sr_2RuO_4$ presents some intriguing superconducting properties. For example, temperature power laws were observed in several transport experiments measuring the attenuation of phonons by electrons, the electronic thermal conductivity and the electronic heat specific heat.

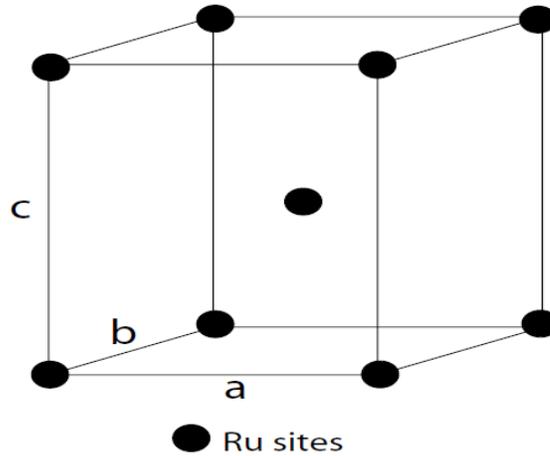

Fig. 1. The body centered tetragonal structure corresponding to an elementary cell of $Sr_2RuO_4$. Both vectors **a** and **b** are of equal magnitude, the ruthenium atoms are shown at the sites of the 3D lattice.

From the beginning, it was proposed that $Sr_2RuO_4$ is an unconventional superconductor (Maeno *et al.*, 1994; Rice and Sigrist, 1995; Miyake and Narikiyo, 1999; Maeno *et al.*, 2001; Zhitomirsky and Rice 2001; Wysokinski *et al.*, 2003; Liu and Mao, 2015; Yarzhemsky, 2018 among many others) with some type of nodes in the order parameter.

Additionally, a series of theoretical and experimental works proposed the existence of a superconducting gap in the triplet state (Ishida *et al.*, 1998; Luke *et al.*, 1998; Duffy *et al.*, 2000; Tanatar *et al.*, 2001 and others) with linear or point nodes, or a combination of them (it was called accidental nodes) in different sheets of its Fermi surface. A triplet superconductor is defined as a superconductor where the electrons pair is bound in spin triplet states.

Moreover, according to several authors, in $Sr_2RuO_4$ the symmetry of the superconducting gap structure is a state of broken time inversion symmetry, with a point symmetry, that transforms as the irreducible two-dimensional representation $E_{2u}$ of the tetragonal point group (see table 2). Although many authors considered that the $\gamma$ sheet of the FS does not have nodes, these works provided experimental agreements with the specific heat $C(T)$, with the electronic heat transport experiments $\kappa_i(T)$, and with the measurements of directional ultrasound $\alpha_i(T)$.

Therefore, all the works indicated in their time, the existence of some type of nodes in addition to the time symmetry breaking state in the pair wave function. Mackenzie and Maeno (2003) provided a literature review of all the works performed during that period of time. In particular, experimental and theoretical ultrasound studies at low temperatures showed the need for an anisotropic electron-phonon interaction in order to fit the experimental results (Lupien *et al.*, 2001; Lupien, 2002; Walker *et al.*, 2001; Wu and Joynt, 2001; Contreras *et al.*, 2004; Nomura, 2005).

This anisotropy in the superconducting state was shown to be inherent to the layered square-lattice structure. Therefore, employing an anisotropic tight-binding model, it was shown (Contreras *et al.*, 2004) that in strontium ruthenate, different "accidental" superconducting nodes correspond to different sheets of the FS.

Using ultrasound fits, it was found for that the $\gamma$ band has 8 point nodes symmetrically distributed in the {100} planes, and 8 point nodes symmetrically distributed in the {110} planes. For the $\alpha$ and/or $\beta$ bands, were found 8 point nodes symmetrically distributed in the {100} planes, and 8 point nodes symmetrically distributed in the {110} planes. Also, it was found that the nodal structure of the $\alpha/\beta$ sheets is different from the **$\gamma$** sheet. The gap on a line along the Fermi surface for **$\alpha/\beta$** sheets is an order of magnitude smaller than that on the **$\gamma$** sheet (see Figure 2).

Furthermore, the exceptionally strong anisotropy in the attenuation of certain modes, which is unique to this material, allowed us to associate the attenuation of the most strongly attenuated modes with their interaction with electrons in the **$\gamma$** sheet. Finally, we found that this "accidental" nodal model satisfies the two-dimensional $E$ irrepresentation for the triplet state superconducting gap (see Table 2 and Figure 2).

In the normal state of $Sr_2RuO_4$, it was shown that to explain the weakness in the attenuation present in the $\gamma$ sheet of the FS (Lupien *et al.*, 2001), the same model yielded the necessary results, since the transverse elastic waves T [100] do not stretch the nearest neighboring bonds, however, they do stretch the second bonds between neighboring atoms (Walker *et al.*, 2001).

In recent years, new experimental studies were performed. They took into account the uniaxial elastic tension in order to determine the gap, and also the time symmetry broken state (Hicks *et al.*, 2014; Taniguchi *et al.*, 2015; Steppke *et al.*, 2017). In that direction, it is possible to point out two novel studies of the jump at the transition temperature to the superconducting state ($T_c$) in the elastic constant $C_{66}$ (Lupien, 2002; Benhabib *et al.*, 2020; and Ghosh *et al.*, 2020).

They studied the transverse elastic propagation velocity polarized in the direction [010] which expanded the studies about the behavior of the sound velocity at the normal-to-superconducting phase previously developed, see also in (Benhabib *et al.* 2020. Jump in the $c_{66}$ shear modulus at the superconducting transition of $Sr_2RuO_4$: Evidence for a two-component order parameter. [arXiv:2002.05916v1](arXiv:2002.05916v1)).

To complete the picture, Table 2 lists a theoretical classification of the in-plane and body centered basis functions for a $D_4$ lattice. I use for the description the group $D_4$ instead of the group $D_{4h}$, this is possible because the **$d$**-vector is an odd function of momentum and the inversion symmetry is already contained in its general form (Contreras, 2006). The irreducible representation $E$ is the one that accounts for the triplet state that breaks the time inversion symmetry in this compound observed in ultrasound studies experiments (Lupien, 2002; Benhabib *et al.* 2020).

In the same order of ideas, and in order to account for the time symmetry broken state, it was necessary to propose Ehrenfest relationships capable of predicting the jump in the elastic tensor $C_{66}$ observed experimentally (Lupien, 2002) and that took into account the role of the transverse elastic tension using the Ginzburg-Landau theory (Walker and Contreras, 2002; Sigrist, 2002; Contreras *et al.*, 2016). These studies allowed the construction of a phase diagram of the compound based on three states: the normal state, the BCS-type singlet (one component) state and a triplet state (two components), see table 1 and Figure 3 for more details.

It is necessary to clarify that to reach the singlet state experimentally, elastic stress different from zero must be applied, capable of splitting the transition temperature into two $T_c$, according to the Ehrenfest diagram shown in Figure 3, since the effect of an external uniaxial stress on the basal plane of $Sr_2RuO_4$ is to break the tetragonal symmetry of the crystal. As a consequence of this, when a second order phase transition to the superconducting state occurs, it splits into two transitions. In this way, the 3 quantities calculated using the modified Ehrenfest relations of Table 1, can be described in a phase diagram. Experimentally this condition for elastic uniaxial stress is given by $|\sigma_6(T)| << 1$ in agreement with the phase diagram shown in Figure 3.

For example, an additional result that is hardly discussed in the literature is that the resulting thermal expansion due to this type of breaking symmetry is zero for zero elastic uniaxial stress (Walker and Contreras, 2002). This is shown in Figure 4. Therefore, for this part, we conclude that the experimental and theoretical studies mentioned above have shown that strontium ruthenate is a compound in which an elastic external uniaxial field splits $T_c$, thanks to the interaction of the transverse elastic waves with the Cooper pairs in a triplet state (Figure 3), this leads to a gap corresponding to a time reversal symmetry broken state in the $\gamma$ sheet of the FS.

Additionally, the study of ultrasound attenuation allowed visualizing the position of the nodes of the superconducting gap using an "accidental" nodes model shown in Figure 2. These point nodes are "accidental" in the sense that they are not required by symmetry but exist only if the material parameters have values in a certain range. Also, these point nodes will degenerate into the line nodes discussed by previous authors. This interpretation led to the determination of a superconducting gap model with different nodal structures on different bands, based on directional ultrasound experiments (Contreras, Walker and Samokhin, 2004).

Table 1. Types of Ehrenfest relations in crystals of cubic (P) or higher anisotropy ($\sigma_i$).

| Ehrenfest relations | Hydrostatic pressure $\sigma_i = -P$ | Anisotropic stress $\sigma_i$ (Voigt notation) |
|---|---|---|
| thermal expansion $\alpha$ jump | $\Delta\alpha = \Delta C_P \left. \left( d \ln T_c(P) \right) \middle/ d\,P \right.$ | --- |
| thermal stress $\alpha_t$ jump | --- | $\Delta\alpha_i = -\Delta C_\sigma \left. \left( d \ln T_c(\sigma_i) \right) \middle/ d\,\sigma_i \right.$ |
| compressibility K jump | $\Delta K = \Delta\alpha \left( \left. d \ln T_c(P) \middle/ d\,P \right) \right.$ | --- |
| elastic compliances $S_{ij}$ jump | --- | $\Delta S_{ij} = -\Delta\alpha_i \left. \left( d \ln T_c(\sigma_j) \right) \middle/ d\,\sigma_j \right.$ |

Table 2. The irreducible basis functions in a tight binding model for the $z$ component of a triplet order parameter in a $D_4$ tetragonal lattice (Contreras, 2006).

| Irrepresentation | In plane basis functions | Body-centered basis functions |
|:---:|:---:|:---:|
| $A_1$ | × | $\cos\left(\dfrac{k_x a}{2}\right)\cos\left(\dfrac{k_y b}{2}\right)\sin\left(\dfrac{k_z c}{2}\right)$ |
| $A_2$ | × | × |
| $B_1$ | × | × |
| $B_2$ | × | $\sin\left(\dfrac{k_x a}{2}\right)\cos\left(\dfrac{k_y b}{2}\right)\sin\left(\dfrac{k_z c}{2}\right)$ |
| $E$ | $E_{x1} = \sin(k_x a)$ $E_{y1} = \sin(k_y a)$ | $E_{x2} = \sin\left(\dfrac{k_x a}{2}\right)\cos\left(\dfrac{k_y b}{2}\right)\cos\left(\dfrac{k_z c}{2}\right)$ $E_{y2} = \cos\left(\dfrac{k_x a}{2}\right)\sin\left(\dfrac{k_y b}{2}\right)\cos\left(\dfrac{k_z c}{2}\right)$ |

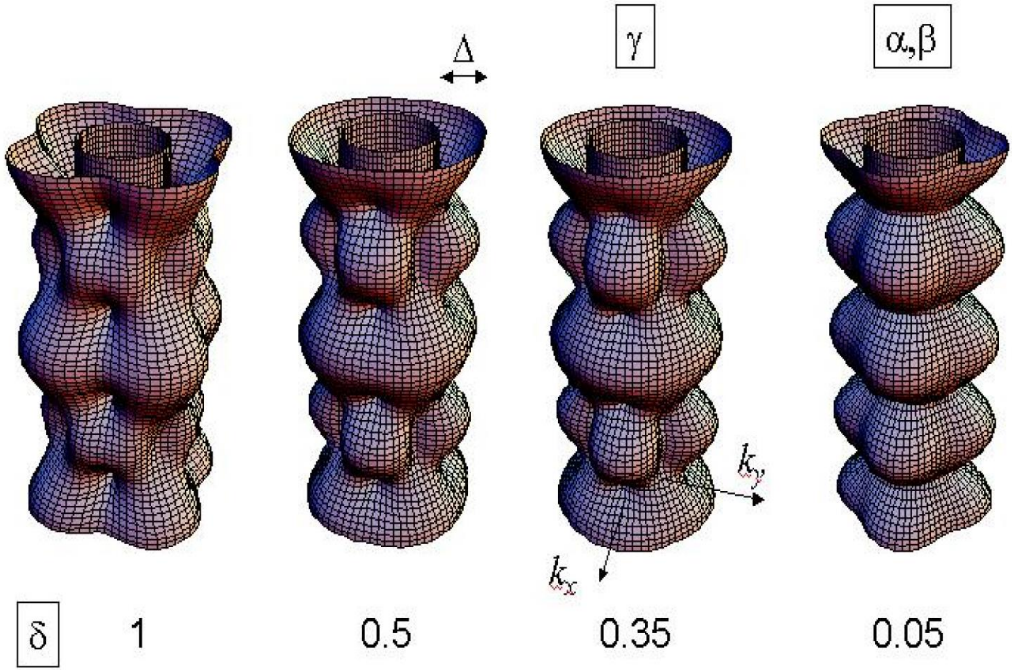

Fig. 2. The anisotropic superconducting triplet gap for Sr$_2$RuO$_4$, according to the model with accidental nodes, and which belongs to a gap with E$_{2u}$ body centered basis functions of the point group D$_{4h}$ in a tetragonal crystal (Contreras et al., 2004).

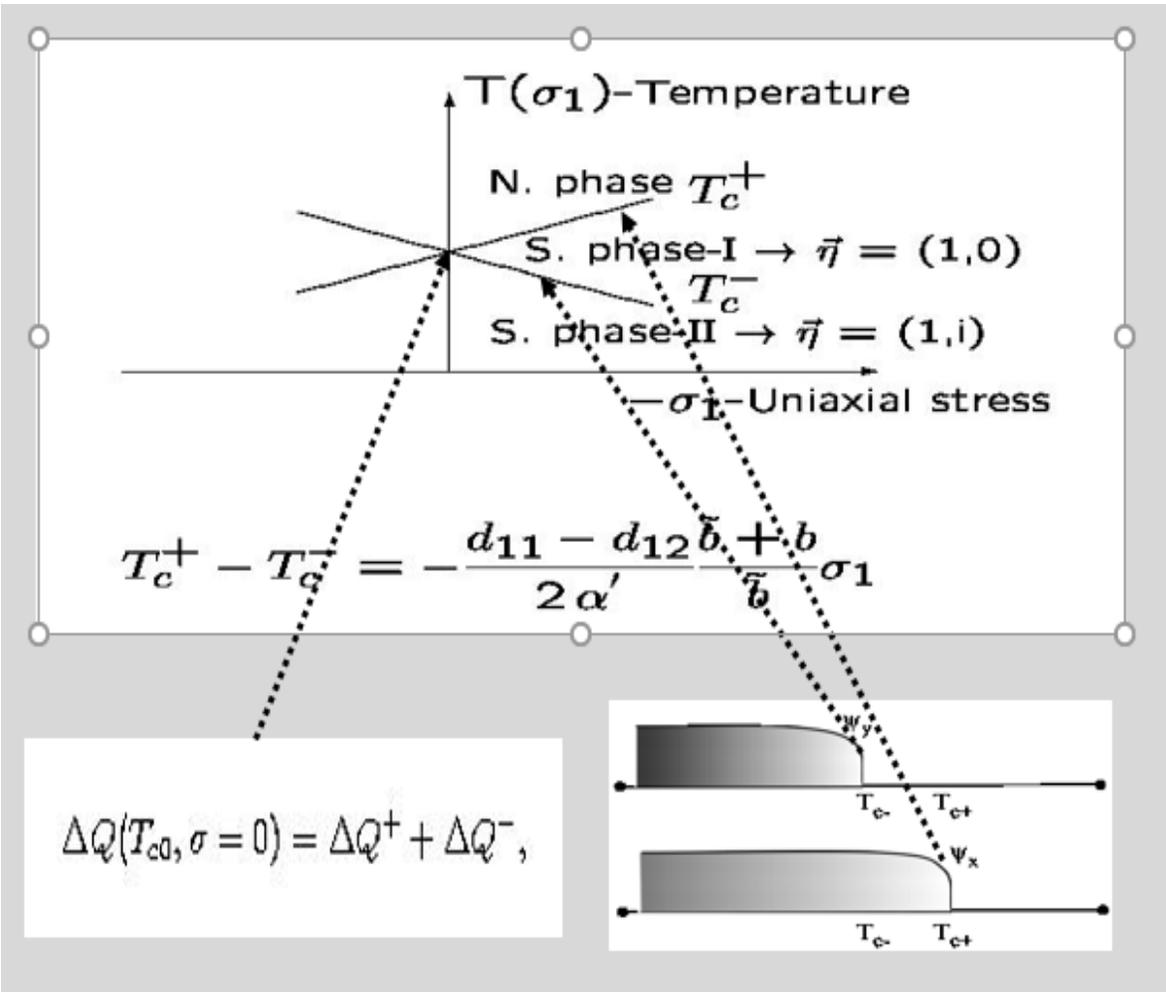

Fig. 3. The phase diagram according to the Ehrenfest relationships for elastic stress at temperatures $T_c$, 0, $T_{c+}$, and $T_{c-}$ for a tetragonal crystal (Contreras, 2006).

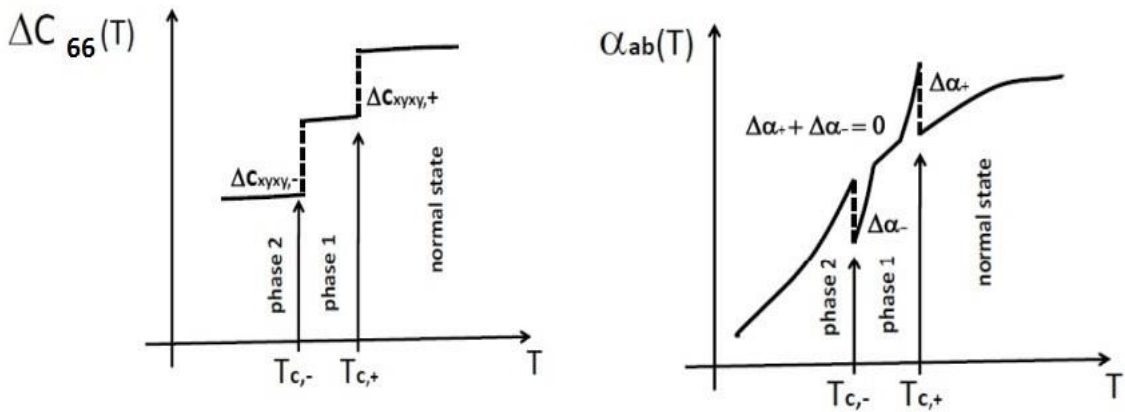

Fig. 4. The jumps the elastic constants $C_{66}$ and in the thermal expansion $\alpha_i(T)$ at the temperatures $T_{c+}$ and $T_{c-}$ for an anisotropic crystal of $D_{4h}$ symmetry. Note that the model predicts zero thermal expansion at the temperature $T_c$ (Figure from Contreras and Moreno, 2019).

### 3. ADDITIONAL REMARKS ON THE PHONON TRANSVERSAL ELASTIC FIELD.

In this note, it is proposed to consider additional theoretical studies in the normal state and at $T_c$ of $Sr_2RuO_4$ aiming at expanding the knowledge about the role of threshold phenomena such as the polarization of the transverse elastic waves and their interaction with the conduction electrons in this multiband superconductor. These studies would be mostly related to the polarization and the quantization of the z-value of the spin of the transverse elastic field in body centered tetragonal crystals of the type of Figure 1 and the study of the phonon spectrum at the different gap singularities in this multiband superconductor.

I start by mentioning the need for different approximations of the electron-phonon interaction in different sheets of the Fermi surface in order to fit experimental data, it is clear that the anisotropic electron-phonon interaction among these quasiparticles plays a fundamental role in this compound. Theoretically, it is known for example that only the second neighbors describe the electron-phonon interaction in the transverse direction [010] for elastic waves propagating along [100] direction for the normal and the superconducting states (Walker et al. 2001, Contreras et al. 2004).

Additionally, in $Sr_2RuO_4$ the stiffness $C_{66}$ is related to the speed of the elastic transverse wave $v_s$ [100] with polarization direction [010] (Lupien, 2002; Paglione *et al*., 2002). Herewith, we can conclude that $Sr_2RuO_4$ is a strong candidate to study microscopic phenomena inherent to transversal elastic waves propagation such as the transversal phonon spin z-value and polarization in anisotropic normal state metallic compounds. However, in order to extend any microscopic study of transverse phonons, a second quantization mean field study of the elastic wave field effects is necessary.

This theoretical program was carried out by Vonsovskii and Svirskii (1961) and Levine (1962), see also discussion in (Kittel, 1958). Vonsovskii and Svirskii in their seminal paper, demonstrated that for transverse elastic waves in isotropic elastic media, the z projection of the expected phonon mean value is equal to ± 1 (depending on the polarization of the wave), unlike longitudinal waves which have zero spin mean z-value projection.

Therefore, a second quantization program of the elastic field in a tetragonal lattice based on the formalism of the phonon quasiparticles creation and annihilation operators might show that the spin of the transverse waves has an integer value of ± 1. For this study, we could use the Lagrangian $\mathcal{L}$ obtained by subtracting the Helmholtz free energy $F$ (Landau and Lifshitz, 1970) from the kinetic energy

$$\mathcal{L} = \frac{1}{2}\rho\,\dot{u}^2 - [\frac{1}{2}\,c_{11}\,(u_1^2 + u_2^2) + \frac{1}{2}\,c_{33}\,u_3^2 + c_{13}(u_1u_3 + u_2u_3) + c_{12}(u_1 + u_2) + 2\,c_{66}\,u_6^2 + 2\,c_{44}\,(u_4^2 + u_5^2)]$$

The Lagrangian for a tetragonal crystal in the Voight notation has six elastic constants $c_{ij}$ and also six invariants for displacement combinations $u_i\,u_j$. The quantization of the elastic field within the previous expression can be obtained in principle with the used of the Noether theorem and the energy impulse tensor as done for the isotropic longitudinal and transversal elastic modes (Vonsovskii and Svirskii, 1961). The mentioned authors accomplished successfully the calculation of the spin tensor $S_{k,lm}$ and the spin density $s$ for an isotropic elastic tensor using the second quantization formalism (Bogoliubov and Shirkov, 1959). for longitudinal and transverse phonon quasiparticles.

In particular, Vonsovskii and Svirskii pointed out in 1961 that for elastic isotropic waves the spin component projection $S_3$ can be theoretically calculated using the expression

$$S_3 = 2 \, i \, \rho \int u_\tau^+ \times u_\tau^- \, dk$$

Where $u_\tau^{+/-}$ describe the transversal elastic field displacement for +/- frequencies in Fourier components. Using the second quantization and the respective diagonalization of the field, Vonsovskii and Svirskii showed that longitudinal phonons in an isotropic crystal have an energy $\hbar\omega_l$ and a mean spin value $< S_3 >= 0$ , meanwhile for transverse phonons, the energy is given by $\hbar\omega_\tau$ and spin value $< S_3 >= \pm \, \hbar$.

Furthermore, in Sr$_2$RuO$_4$ there are experimental studies with the observation of the breaking of the inversion symmetry due to the change in velocity of the elastic transverse wave that propagates in the propagates in the [100] direction with [010] polarization at $T_c$. Henceforth, the possibility for the quantization of the transverse phonon field in terms of creation and annihilation operators might bring a link to the elastic field symmetry-breaking effect in this ternary compound and additionally to other threshold phenomena such that the change in the phonon dispersion law at the gaps value.

Additionally, in another order of ideas, a new theory has recently been formulated by Grechka (2020) where the classic Christoffel equation (Musgrave, 1970) is solved in the polarization variables. This finding makes it possible to study the polarization fields due to the propagation of elastic waves for homogeneous anisotropic crystals. Solving for the slowness vector $p(U)$ (the inverse of the phase velocity) of the plane waves corresponding to a given polarization, unexpectedly, Grechka found a subset of triclinic solids in which the polarization field contains holes; showing that there are solid angles of finite size with polarization directions unattainable for any plane wave, depending on the value of the elastic constants.

Henceforth, this remarkable work shows that the study of the polarization for elastic fields in tetragonal media such as Sr$_2$RuO$_4$ can be of interest. In particular, in the γ sheet, ballistic phonon studies, where the electron-phonon interaction is weak and is coupled to the transversal mode. For studies in niobium see (Hauser, Gaitskell and Wolfe, 1999). Table 3 summarizes the different pathways to solve the Christoffel equation in anisotropic crystals (Grechka 2017, Grechka 2020).

Table 3. Solution methods for the Christoffel equation in anisotropic crystals according to Grechka (2017, 2020)

| Eigenvalues of the Christoffel matrix for phase velocities $\Gamma(n)$, as function of the wave front normal $n$. | Eigenvalues of the Christoffel matrix $\Gamma(p)$ for phase slowness $p$. | Eigenvalues of the Christoffel matrix $\Gamma(U)$ for phase slowness $p(U)$ as a function of the polarization vector $U$. |
|---|---|---|

Finally, a theoretical microscopic study of the change in speed of the transverse elastic waves in a system with tetragonal symmetry such as Sr$_2$RuO$_4$ at $T_c$ could be of importance. For example, the calculation of the real part of the phonon polarization operator $\Re\,\Pi_i^R(\omega, q)$ (see figure 5) would potentially elucidate, if there are singularities in the energy spectrum of the transverse elastic phonon field at $T_c$. For studies on other superconducting systems, see (Kulik, 1963; Vaskin and Demikhovskii, 1968). The real part of the phonon polarization $\Pi_i^R(\omega, q)$ is given according to

$$\Re\,\Pi_i^R(\omega, q) = \frac{1}{\pi} \int_0^\infty \Im\,\Pi_i^R(\omega, q) \left( \frac{1}{x - \omega} + \frac{1}{x + \omega} \right) dx$$

where $\Im\,\Pi_i^R(\omega, q)$ is the imaginary part of the polarization operator and $\Re\,\Pi_i^R(\omega, q)$ is the real part (Vaskin and Demikhovskii, 1968).

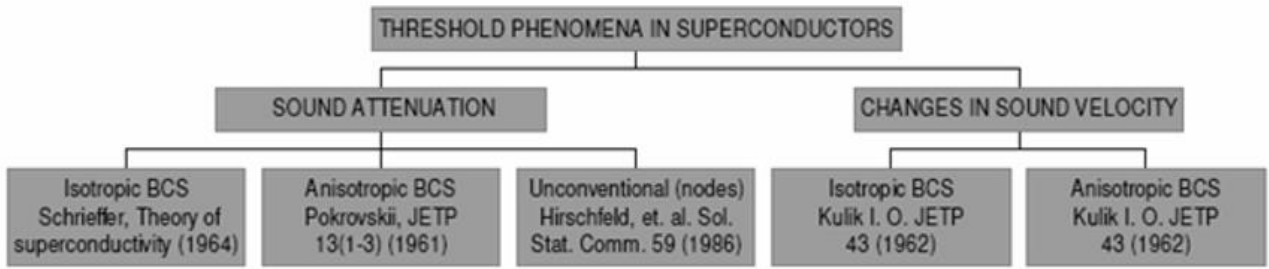

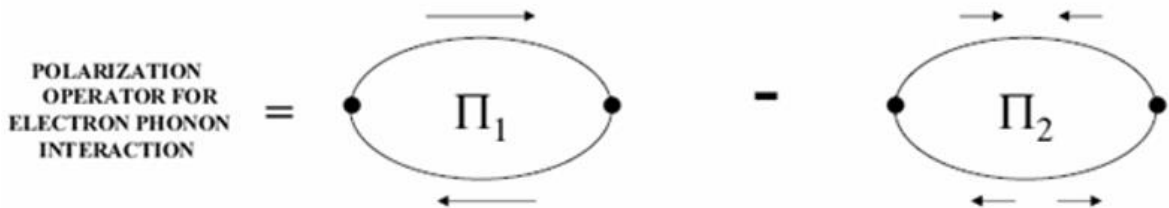

Fig. 5. Some superconducting theoretical threshold studies from the 60s and 80s of the Green functions polarization operator $\boldsymbol{\pi}$ for the electron-phonon interaction and the phonon spectrum in certain classes of superconductors.

## 4. CONCLUSION

The unconventional superconductivity in the multiband superconductor $Sr_2RuO_4$ was of great interest for theoreticians and experimentalists during the last century. It is obvious that this century researches will continue to focus on this phenomenon. Therefore, this short discussion has review some aspects of the problem related to several threshold phenomena, that it might be worth exploring in this compound.

In particular, the present work was aimed at investigating some literature of those threshold elastic phenomena at $T_c$ where the dispersion laws and the spin nature for the quasiparticle phonons (longitudinal and transverse) and conduction electrons might play a relevant role to help explain different physical properties of this multiband compound.

## 5. ACKNOWLEDGEMENTS

This research did not receive any grant from the University of Los Andes or national government agency. P. Contreras acknowledges a previous version published thanks to the effort of the Canadian Journal of Pure and Applied Sciences by helping the most disadvantaged scientists around the world.

## 6. REFERENCES

Ambegaokar, V. and Griffin, A. 1965. Theory of the Thermal Conductivity of Superconducting Alloys with Paramagnetic Impurities. Phys. Rev. 137, A1151.

DOI: https://doi.org/10.1103/PhysRev.137.A1151

Benhabib, S., Lupien, C., Paul, I., Berges, L., Dion, M., Nardone, M., Zitouni, A., Mao, ZQ., Maeno, Y., Georges, A., Taillefer, L. and Proust, C. 2020. Ultrasound evidence for a two-component superconducting order parameter in $Sr_2RuO_4$. Nature Physics. 6 pages. DOI: https://doi.org/10.1038/s41567-020-1033-3

Bergemann, C., Mackenzie, A., Julian, S. J., Forsythe D. & Ohmichi E. 2003 Quasi-two-dimensional Fermi liquid properties of the unconventional superconductor Sr2RuO4, Advances in Physics, 52:7, 639-725, DOI: 10.1080/00018730310001621737

Bogoliubov, N. and Shirkov D, 1959. Introduction to the theory of quantized fields, John Wiley & Sons, New York.

Contreras, P., Walker, MB. and Samokhin, K. 2004. Determining the superconducting gap structure in $Sr_2RuO_4$ from sound attenuation studies below $T_c$. Physical Review B. 70(18):184528. DOI: https://doi.org/10.1103/PhysRevB.70.184528

Contreras, P. Walker, MB. and Samokhin K. 2004. Determining the superconducting gap structure in Sr2RuO4 from sound attenuation studies below Tc. Phys. Rev. B, 70:184528, 2004. doi:10.1103/PhysRevB.70.184528

Contreras, PL. 2006. Symmetry properties and sound waves in the unconventional superconductor $Sr_2RuO_4$. Ph.D. Thesis. University of Toronto, Canada.

Contreras, P. 2011. Electronic heat transport for a multiband superconducting gap in $Sr_2RuO_4$. Revista Mexicana de Física. 57(5):395-399.

Contreras, P., Flórez, J. and Almeida, R. 2016. Symmetry field breaking effects in $Sr_2RuO_4$. Revista Mexicana de Física. 62(5):442-449.

Contreras, P. and Moreno, J. 2019. Anisotropic shear stress $\sigma_{xy}$ effects in the basal plane of $Sr_2RuO_4$. Canadian Journal of Pure and Applied Sciences. 13(2):4807-4812.

Duffy, JA., Hayden, SM., Maeno, Y., Mao, Z., Kulda, J. and McIntyre, GJ. 2000. Polarized-neutron scattering study of the Cooper-pair moment in $Sr_2RuO_4$. Physical Review Letters. 85(25):5412-5415. DOI: https://doi.org/10.1103/PhysRevLett.85.5412

Ghosh, S., Shekhter, A., Jerzembeck, F., Kikugawa, N., Sokolov, DA., Brando, M., Mackenzie, AP., Hicks, CW. and Ramshaw, BJ. 2020. Thermodynamic evidence for a two-component superconducting order parameter in $Sr_2RuO_4$. Nature Physics. 9 pages. DOI: https://doi.org/10.1038/s41567-020-1032-4

Grechka, V. 2020. Christoffel equation in the polarization variables. Geophysics. 85(3):C91. DOI: https://doi.org/10.1190/geo2019-0514.1

Grechka, V. 2017. Algebraic degree of a general group-velocity surface. Geophysics 82: WA45-WA53. https://doi.org/10.1190/geo2016-0523.1

Griffin, A. 1965. The electronic thermal conductivity and other properties of gapless superconductors. Ph.D. Thesis. Cornell University, USA.

Hauser, M., R. Gaitskell, R., and Wolfe, J. 1999. Imaging phonons in a superconductor. Phys. Rev. B 60, 3072. DOI: https://doi.org/10.1103/PhysRevB.60.3072

Hicks, CW., Brodsky, DO., Yelland, EA., Gibbs, AS., Bruin, JAN., Barber, ME., Edkins, SD., Nishimura, K., Yonezawa, S., Maeno, Y. and Mackenzie, AP. 2014. Strong increase of $T_c$ of $Sr_2RuO_4$ under both tensile and compressive strain. Science. 344(6181):283-285. DOI: https://doi.org/10.1126/science.1248292

Hirschfeld, PJ., Wölfle, P. and Einzel, D. 1988. Consequences of resonant impurity scattering in anisotropic superconductors: Thermal and spin relaxation properties. Physical Review B. 37(1):83. DOI: https://doi.org/10.1103/PhysRevB.37.83

Ishida, K., Mukuda, H., Kitaoka, Y., Asayama, K., Mao, ZQ., Mori, Y. and Maeno, Y. 1998. Spin-triplet superconductivity in $Sr_2RuO_4$ identified by [17]O Knight shift. Nature (London). 396:658-660.

Kittel, C. 1958. Interaction of spin waves and ultrasonic waves in ferromagnetic crystals. Physical Review. 110(4):836-841. https://link.aps.org/doi/10.1103/PhysRev.110.836

Kulik, IO. 1963. Heat anomaly of superconductors. Soviet Physics Journal of Experimental and theoretical Physics (JETP, Moscow). 16(4):1952-1954.

Landau, LD. and Lifshitz, EM. 1980. Statistical Physics. Butterworth-Heinemann, Oxford.

Landau, LD. and Lifshitz, EM. 1970. Theory of Elasticity. Pergamon Press, Toronto.

Levine, AT. 1962. A note concerning the spin of the phonon. Nuovo Cimento. 26:190-193. https://doi.org/10.1007/BF02754355

Lifshitz, EM. and Pitaevskii, L. 1987. Kinetical Physics. Butterworth-Heinemann, Oxford.

Liu, Y. and Mao, ZQ. 2015. Unconventional superconductivity in $Sr_2RuO_4$. Physica C: Superconductivity and its Applications. 514:339-353; DOI: https://doi.org/10.1016/j.physc.2015.02.039

Luke, GM., Fudamoto, Y., Kojima, KM., Larkin, MI., Merrin, J., Nachumi, B., Uemura, YJ., Maeno, Y., Mao, ZQ., Mori, Y., Nakamura, H. and Sigrist, M. 1998. Time-reversal symmetry-breaking superconductivity in $Sr_2RuO_4$. Nature (London). 394(6693):558-561. DOI: https://doi.org/10.1038/29038.

Lupien, C. 2002. Ultrasound attenuation in the unconventional superconductor $Sr_2RuO_4$. Ph.D. Thesis. University of Toronto, Canada.

Lupien, C., MacFarlane, WA., Proust, C., Taillefer, L., Mao, ZQ. and Maeno, Y. 2001. Ultrasound attenuation in $Sr_2RuO_4$: An angle-resolved study of the superconducting gap function. Physical Review Letters. 86(26):5986-5989. DOI: https://doi.org/10.1103/PhysRevLett.86.5986

Mackenzie, AP. and Maeno, Y. 2003. The superconductivity of $Sr_2RuO_4$ and the physics of spin-triplet pairing. Reviews of Modern Physics. 75(2):657-712.

Maeno, Y., Hashimoto, H., Yoshida, K., Nishizaki, S., Fujita, T., Bednorz, JG. and Lichtenberg, F. 1994. Superconductivity in a layered perovskite without copper. Nature (London). 372:532-534. DOI: https://doi.org/10.1038/372532a0

Maeno, Y., Rice, TM. and Sigrist, M. 2001. The intriguing superconductivity of strontium ruthenate. Physics Today. 54(1):42-47. DOI: https://doi.org/10.1063/1.1349611.

Miyake, K and Narikiyo, O. 1999. Model for Unconventional Superconductivity of $Sr_2RuO_4$, Effect of Impurity Scattering on Time-Reversal Breaking Triplet Pairing with a Tiny Gap. Phys. Rev. Lett. 83, 1423. DOI: https://doi.org/10.1103/PhysRevLett.83.1423

Musgrave, MJ. 1970. Crystal Acoustics. Holden-Day, San-Francisco, USA.

Nomura, T. 2005. Theory of transport properties in the p-wave superconducting state of $Sr_2RuO_4$ - a microscopic determination of the gap structure. Journal of the Physical Society of Japan. 74(6):1818-1829. DOI: https://doi.org/10.1143/jpsj.74.1818

Paglione, J., Lupien, C., MacFarlane, W., Perz, J., Taillefer, L., Mao, Q. and Maeno, Y. 2002. Elastic tensor of $Sr_2RuO_4$. Physical Review B. 65(22):220506(R). DOI: https://doi.org/10.1103/PhysRevB.65.220506

Pokrovskii, VL. and Toponogov, VA. 1961. Reconstruction of the energy gap in a superconductor by measurement of sound attenuation. Soviet Physics Journal of Experimental and Theoretical Physics (JETP) 13(4):785-786.

Pokrovskii, VL. 1961. Threshold Phenomena in Superconductors. Soviet Physics Journal of Experimental and Theoretical Physics (JETP). 13(1):100-104

Rice, TM. and Sigrist, M. 1995. $Sr_2RuO_4$: an electronic analogue of $^3He$? Journal of Physics: Condensed Matter. 7(47):L643-L648.

Rumer M. and Ryvkin Yu. 1980, Thermodynamics, Statistical Physics, and Kinetics, Mir publishers.

Sigrist, M. 2002. Ehrenfest relations for ultrasound absorption in $Sr_2RuO_4$. Progress of Theoretical Physics. 107(5):917-925. DOI: https://doi.org/10.1143/PTP.107.917

Steppke, A., Zhao, L., Barber, ME., Scaffidi, T., Jerzembeck, F., Rosner, H., Gibbs, AS., Maeno, Y., Simon, SH., Mackenzie, AP. and Hicks, CW. 2017. Strong peak in $T_c$ of $Sr_2RuO_4$ under uniaxial pressure. Science. 355(6321):eaaf9398. DOI: https://doi.org/10.1126/science.aaf9398

Taniguchi, H., Nishimura, K., Goh, SK., Yonezawa, S. and Maeno, Y. 2015. Higher-$T_c$ superconducting phase in $Sr_2RuO_4$ induced by in-plane uniaxial pressure. Journal of the Physical Society of Japan. 84(1):014707. DOI: https://doi.org/10.7566/JPSJ.84.014707

Tanatar, MA., Nagai, S., Mao, ZQ., Maeno, Y. and Ishiguro, T. 2001. Thermal conductivity of superconducting $Sr_2RuO_4$ in oriented magnetic fields. Physical Review B. 63(6):064505. DOI: https://doi.org/10.1103/PhysRevB.63.064505

Vaskin, V. and Demikhovskii, V. 1968. Sound dispersion in superconducting semiconductors. Soviet Physics of the Solid State. 10(2):330-333.

Vonsovskii, SV. and Svirskii, MS. 1961. About the spin of phonons. Soviet Physics of the Solid State. 3:2160. In Russian.

Walker, MB. 1980. Phenomenological theory of the spin-density-wave state of chromium. Physical Review B. 22(3):1338-1347. DOI: https://doi.org/10.1103/PhysRevB.22.1338

Walker, MB., Smith, M. and Samokhin, K. 2001. Electron-phonon interaction and ultrasonic attenuation in the ruthenate and cuprate superconductors. Physical Review B. 65(1):014517. DOI: https://doi.org/10.1103/PhysRevB.65.014517

Walker, MB. and Contreras, P. 2002. Theory of elastic properties of $Sr_2RuO_4$ at the superconducting transition temperature. Physical Review B. 66(21):214508. DOI: https://doi.org/10.1103/PhysRevB.66.214508

Wu, WC. and Joynt, R. 2001. Transport and the order parameter of superconducting $Sr_2RuO_4$. Physical Review B. 64(10):100507(R). DOI: https://doi.org/10.1103/PhysRevB.64.100507

Wysokiński, KI., Litak, G., Annett, JF. and Györffy, BL. 2003. Spin triplet superconductivity in $Sr_2RuO_4$. Physica Status Solidi (b). 236(2):325-331. DOI: https://doi.org/10.1002/pssb.200301672

Yarzhemsky, VG. 2018. Group theoretical lines of nodes in triplet chiral superconductor $Sr_2RuO_4$. Journal of the Physical Society of Japan. 87(11):114711. DOI: https://doi.org/10.7566/JPSJ.87.114711

Zhitomirsky, M. and Rice, T. 2001. Interband proximity effect and nodes of superconducting gap in $Sr_2RuO_4$. Phys. Rev. Lett. 87, 057001. DOI: https://doi.org/10.1103/PhysRevLett.87.057001